\documentstyle[12pt]{article}
\newcommand{\R}{{\bf R}}
\newcommand{\be}[1]{\begin{equation}\label{#1}}
\newcommand{\ee}{\end{equation}}
\newcommand{\beq}[1]{\begin{eqnarray}\label{#1}}
\newcommand{\eeq}{\end{eqnarray}}
\newcommand{\beqo}{\begin{eqnarray*}}
\newcommand{\eeqo}{\end{eqnarray*}}
\newcommand{\li}[1]{\left#1}
\newcommand{\re}[1]{\right#1}

\newcommand{\half}{\frac{1}{2}}

\newcommand{\suml}{\sum\limits}
\newcommand{\qmbox}[1]{\quad\mbox{#1}\quad}

\newcommand{\ga}{\alpha}
\newcommand{\gb}{\beta}
\newcommand{\ba}{\begin{array}}
\newcommand{\ea}{\end{array}}
\newcommand{\dddot}[1]{\ddot{#1}\dot{\hspace{-1.4ex}#1}}

\author{Sabine Kluske and Hans - J\"urgen Schmidt}
\title{Towards a no hair theorem for higher order gravity}
\date{}
\begin{document}
\maketitle

\bigskip

\centerline{
Universit\"at  Potsdam, Mathematisch-naturwiss. Fakult\"at }
\centerline{  Projektgruppe Kosmologie }
\centerline{  D-14415 POTSDAM, PF 601553, Am Neuen Palais 10,
Germany }
\centerline{  e-mail hjschmi@rz.uni-potsdam.de,
kluske@rz.uni-potsdam.de }

\bigskip
\begin{abstract}
We use   gravitational lagrangians $
R  \Box \sp k R \sqrt{-g}$ and linear combinations of them; we
ask under
which circumstances the de Sitter space-time represents an
attractor solution
in the set of spatially flat Friedman models.

Results are: For arbitrary  $k$, i.e., for arbitrarily large
order $2k+4$ of the
field equation, on can always find examples where the
attractor property
takes place. Such examples necessarily need a non-vanishing
$R\sp 2$-term.
The main formulas  do not depend on the dimension, so one gets
similar
results also for 1+1-dimensional gravity and for Kaluza-Klein
cosmology.

\end{abstract}

of gravitation

\bigskip

\section{ Introduction }

Over the years, the notion ''no hair conjecture'' drifted to
''no hair theorem''
without possessing a generally accepted formulation  or even a
complete proof.
Several trials have been made to formulate and prove it at
least for certain
special cases. They all have the overall structure: ''For a
geometrically
defined class of space-times and physically motivated
properties of the
energy-momentum tensor, all the solutions of the gravitational
field equation
tend asymptotically to a space of constant curvature.''

\subsection{Historical notes}

It is the aim of this paper to clarify the relations between
the several
existing versions, and then to develop the cosmological no
hair theorem
towards applicability to a certain class of higher order field
equations.
Let us start with some historical notes.

The paper [1] by Weyl (1927) is cited in [2] with the phrase
''The
behaviour of every world satisfying certain natural
homogeneity conditions
in the large follows the de Sitter solution asymptotically.''
to be the
first published version of the no hair conjecture.

Barrow and G\"otz [2] apply the formulation ''All
ever-expanding universes
with $\Lambda >0$ approach the de Sitter space-time locally.''

 (Ever--expanding to be meant as: there is a time $t_0$ such
that for all $t>t_0$ the Hubble parameter is positive. In
other words: a bounce is allowed, a recollapse is not
allowed.)

Let us comment
this formulation: So they circumvent the necessity to
distinguish the initial
data between expanding and recollapsing ones, but their
formulation needs a
further explanation; example: If one changes the initial data
continuously
from recollapsing to ever-expanding ones, then one gets a
critical value of
the initial data between them, where one has an ever-expanding
model which
need not tend to the de Sitter space-time but can have
typically a linear
expansion law. So these critical values of the initial data
have to be
excluded, too.

The first proof of the stability of the de Sitter solution
(here: within the
steady-state theory), is due to Hoyle and Narlikar [3]. In the
papers [4]
by Price perturbations of scalar fields have been
considered for the no hair theorem.
The probability of inflation is large if the no hair theorem
is valid, cf.
[5].

\subsection{Physical properties}

Peter, Polarski and Starobinsky [6] compared the
double-inflationary models
with cosmological {\it observations}.
Barrow and G\"otz discussed the no hair conjecture within {\it
Newtonian}
cosmological models [2,7].
H\"ubner and Ehlers considered {\it inflation in an open
Friedman universe}
and have noted that inflationary models need not to be
spatially flat [8].

Gibbons and Hawking have found two of the earliest strict
results on the
no hair conjecture for Einstein's theory [9] in 1977.
Barrow gave examples that the no hair conjecture fails if the
{\it
energy condition} is relaxed and points out, that this is
necessary to
solve the graceful exit problem. He uses the formulation of
the no hair
conjecture ''in the presence of an effective cosmological
constant (e.g.
from viscosity) the de Sitter space-time is a stable
asymptotic solution''.
This is a much weaker statement because only space-times in a
neighbourhood
of the de Sitter space-time are involved. He mentioned that an
ideal fluid
with equation of state $p=-\rho$ is equivalent to a
$\Lambda$-term
in some cases but not always [10].

Usually, energy inequalities are presumed for formulating the
no hair
conjecture. Nakao, Shiromizu and Maeda [11] found some cases
where it
remains valid also for negative Abbott-Deser mass [the latter
goes over to
the well-known ADM-mass (Arnowitt, Deser, Misner) for $\Lambda
\longrightarrow 0$]. They cite Murphy [12]. In [12], {\it
viscosity
terms} as source are considered to get a singularity-free
cosmological model.
Murphy [12] used Einstein's theory, and Oleak [12] made
similar
considerations within Treder's theory of gravity.

In the Eighties, these non-singular models with viscosity
where
re-inter\-pre\-ted as inflationary ones, cf. e.g. [13].

In the three papers [14] Prigogine et al. developed a
phenomenological model
of particle and {\it entropy creation}. It allows particle
creation
from space-time curvature, but the inverse procedure (i.e.
particle decay into space--time curvature) is forbidden.
This breaks the
$t\longrightarrow -t$-invariance of the model. Within
that model, the expanding de Sitter space-time is an attractor
solution independently of the initial fluctuations; this
means, only the expanding de Sitter
solution is thermodynamically possible. To these papers cf.
also [15].

Vilenkin [16] discussed {\it future-eternal} inflating
universe models;
they must have a singularity if the condition D: ''There is at
least one
point $p$ such that for some point $q$ to the future of $p$
the volume of
the difference of the pasts of $p$ and $q$ is finite'' is
fulfilled.

Mondaini and Vilar have considered {\it recollapse} and the no
hair
conjecture in closed higher-dimensional Friedman models [17].
Pullin [18] discussed relations between the onset of black
hole
formation and the no hair conjecture.
Concerning the no hair conjecture Shiromizu, Nakao, Kodama and
Maeda [19]
gave the following argument: If the matter distribution is too
clumpy, then
a large number of small black holes appears. Then one should
look for an
inflationary scenario where these black holes are harmless.
They cannot
clump together to one giant black hole because of the
exponential expansion
of the universe; this explains the existing upper bound of
black holes in the quasi-de
Sitter  model.
Shibata, Nakao, Nakamura and Maeda have considered asymptotic
gravitational
waves in an axially symmetric quasi de Sitter space-time [20].
They use
numerical methods. The magnitude of black holes is restricted:
above
$M_{crit} = \frac{1}{3\sqrt\Lambda}$ there do not exist
horizons; this
restriction one gets by considering a perturbed
Schwarzschild-de
Sitter-solution. The cosmic {\it hoop conjecture} expresses
that the mass
of a black hole in a quasi de Sitter model is bounded from
above by
$M_{crit} = \frac{1}{3\sqrt {\Lambda}}$, and its surface  is
analogously restricted.

The notion {\it ''quantum hair''} means quantum numbers
presenting quantum fields
which should be classically forbidden if the no hair theorem
is valid.
Coleman, Preskill and Wilczek found examples of quantum hairs
on black holes
[21].

Xu, Li and Liu [22] proved the instability of the {\it anti-de
Sitter}
space-time (classical instability against gravitational waves,
and dust
matter perturbations); one has $\Lambda < 0$, and in an open
Friedman
model the scale factor $a$ in dependence of synchronized time
$t$ reads
$a=\alpha \, \cos \frac{t}{\alpha}$ where $\Lambda = -
3/\alpha \sp 2 $.
The anti-de Sitter model has closed time-like curves
everywhere; a Cauchy
horizon is the surface where closed time-like curves begin to
exist, and
therefore, the anti-de Sitter model has no Cauchy horizon. (Of
course, a closed curve has no beginning; the formulation
means: The Cauchy horizon is the topological boundary of the
set of point possessing the property that they are contained
in a  closed  time--like curve.)

Coley and Tavakol discussed the robustness of the cosmic no
hair conjecture
under using the concept of the structural stability [23]
(compare with
chapter 5 below).

\subsection{Fourth-order gravity}

Sirousse-Zia considered the Bianchi type IX model in
Einstein's theory
with a positive $\Lambda $-term and got an asymptotic
isotropization of
the mixmaster model [24]. She cites (and uses methods of)
Belinsky,
Lifshitz and Khalatnikov [25].
M\"uller [26] used $L=R^2$ and discussed the power-asymptotes
of Bianchi
models.
Barrow and Sirousse-Zia [27] discussed the {\it mixmaster}
$R\sp 2$-model
and the question, under which conditions  the Bianchi type IX

model becomes asymptotic de Sitter ?

Yokoyama and Maeda [28] considered the no hair conjecture for
Bianchi type
IX models and Einstein's theory with a cosmological term. They
discussed
$R\sp 2$ inflation in anisotropic universe models and got as a

result that
typically, an initial anisotropy helps to enhance inflation.
For Bianchi type
IX they got some recollapsing solutions besides those
converging to the
de Sitter solution.

Cotsakis, Demaret and de Rop [29] discussed the mixmaster
universe in
fourth-order gravity. To take the {\it metric diagonal} they
write ''is probably
well justified''; they discuss all types of curvature-squared
terms.
Paper [27] is continued in [30] by Spindel, where also general
Bianchi type I
models in general dimensions are considered.

Gurovich et al. [31] considered $L=R+\alpha R\sp{4/3}$ to get
a
singularity-free model in 1970, the solutions are of a quasi
de Sitter type.
One should remember that in spite of de Sitter's papers in the
twenties,
the inflationary cosmological model became generally accepted
in only
in 1979/80.

The papers [32], [33] consider the {\it no hair conjecture}
for
$R\sp 2$ models, they use the formulation ''asymptotical de
Sitter, at
least on patch''. The restriction ''on patch'' is not strictly
defined but
refers to a kind of local validity of the statement, e.g., in
a region
being covered by one single synchronized system of reference
in which the
spatial curvature is non-positive and the energy conditions
are fulfilled.
The Starobinsky model is outlined as one which does not need
an additional
inflaton field to get the desired quasi de Sitter stage. One
should
observe a notational change: There,
$L=R+ a R\sp 2 \ln R$ was called Starobinsky model, whereas
$L=R+ a R\sp 2 $ got the name ''improved Starobinsky model'' -
but now
the latter carries simply the name ''Starobinsky model''. (For
the
inflationary phase, both versions are quite similar.) A
further result of
the papers [32] is that by the addition of a cosmological
term, the
Starobinsky model leads naturally to double inflation. Let us
comment
this result: It is correct, but one should add that this is
got at the
price of getting a ''graceful exit problem'' (by this phrase
there is ment
the problem of how to finish the inflationary phase
dynamically) - in the
Starobinsky model this problem is automatically solved by the
fact that
the quasi de Sitter phase is a transient attractor only.
The papers [34] discuss the no hair conjecture within $R\sp
2$-models and
found inflation as a transient attractor in fourth order
gravity.
The papers [32] and [35], [37] discuss the stability of
inflation in
$R\sp 2$-gravity.
The papers [36] discuss generalized cosmic no hair theorems
for quasi
exponential expansion.
In Starobinsky [38] the no hair theorem for Einstein's theory
with
a positive $\Lambda$-term is tackled by using a sequence as
ansatz to
describe a general space-time. However, the convergence of the
sequence
is not rigorously proven.

Starobinsky and Schmidt [39] have generalized the ansatz of
Starobinsky
[38] to consider also the no hair theorem for $L=R\sp 2$.

Shiromizu et al. [19] discussed an inflationary inhomogeneous
scenario and
mention the open problem {\it how to define asymptotical de
Sitter}
space-times.
In Pacher [40]  it is mentioned that only a local version of
this
conjecture can be expected to hold true, and that neither the
definition
of asymptotic de Sitter nor the necessary presumptions to the
energy-momentum
tensor are clarified - two problems which are not finally
solved up to now.
The authors of [41] consider the no hair theorem for a special
class
of inhomogeneous models and give partial proofs.
Morris [42] considers inhomogeneous models for $R+R\sp
2$-cosmology.
In [43] inflation in inhomogeneous but spherically symmetric
cosmological
models is obtained only if the Cauchy data are homogeneous
over several
horizon lengths.
The  analogous problem is considered in [44] also with
inclusion of
colliding plane gravitational waves, they give a numerical
support of the no
hair conjecture by concentrating on the dynamics of
gravitational waves.

Berkin [45] gets as further result, that for $L=f(R)$, a
diagonal Bianchi
metric is always possible.
Similarly, Barrow and Sirousse-Zia [27] and Spindel [30]
worked on
{\it diagonalization} problem. They apply the diagonalisation
condition of
MacCallum et al. [46].
In 1918 Kottler [47] found a simple closed-form static
spherically symmetric
vacuum solution for Einstein's theory with $\Lambda$-term in
Schwarzschild
coordinates.
\begin{equation}
ds\sp 2 = A(r) dt\sp 2 - \frac{dr\sp 2}{A(r)}
- r\sp 2 (d\theta \sp 2 + \sin \sp 2 \theta \, d \phi \sp 2 )
\end{equation}
with $A(r)=1 - \frac{2m}{r} - \frac{\Lambda}{3} r\sp 2$.
At the horizon $A=0$ the Killing vector changes its sign and
one
gets by interchanging the coordinates $t$ and $r$ the
corresponding
{\it Kantowski-Sachs} model.
Moniz [48] (1993) discusses the cosmic no hair conjecture
within
Kantowski-Sachs models and $\Lambda > 0$. He gets the de
Sitter space-time
not only asymptotically, but exactly in an anisotropic
3+1-slicing
of space-time. He discusses the initial data that lead to a
recollapse and
find them to be very rare; but the measure he uses is not
well-defined,
so, possibly, this is not the last word. It is curious to
observe that he
works with complicated elliptic integrals instead of applying
the
Schwarzschild-de Sitter-solution found by Kottler [47] in
1918.

[49] gives an overview about the geometry of the de Sitter
space-time.

\subsection{Sixth and higher order models}

The paper [50] by Buchdahl (1951) deals with lagrangians of
arbitrarily high
order. Its results are applied in paper [51] to general
Lagrangians $F(R,
\Box)$.
{}From another motivation, Bollini et al. [52] consider
higher-order field
theories of the type $$\sum a_s \Box \sp s \phi (x) = 0$$
and give solutions in the sense of distributions.

Forgacs et al. [53] consider the non-local lagrangian
$R\frac{1}{\Delta}R-M$
as Wess Zumino Witten model.

The paper [54] by Vilkovisky was presented at the A.
Sacharov-memorial
conference held in Moscow in May 1991. In [54], the
Sacharov-approach
was generalized. The original idea of Sacharov (in 1967) was
to define
higher order curvature corrections to the Einstein action to
get a kind
of elasticity of the vacuum. Then the usual breakdown of
measurements at
the Planck length (such a short de Broglie wave length
corresponds to such
a large mass which makes the measuring apparatus to a black
hole) is
softened. Vilkovisky discusses the effective gravitational
action in the
form $R f(\Box )R$, where
$$f(\Box )= \int \frac{1}{\Box - x } \rho(x)dx$$

Martin and Mazzitelli [55] discuss the non-local Lagrangian
$R\frac{1}{\Box }R$ as conformal anomaly in two dimensions.

Let us now come to {\it sixth--order} equations. Stelle [56]
(1977) considers
mainly fourth order $R\sp 2$-models; in the introduction he
mentioned
that in the next order, terms like $R\sp 3 + R_{ij;k}
R\sp{ij;k}$
become admissible, but  the pure $R\sp 3$-term is not
admissible.
Treder [57] used higher-order lagrangians, erspecially $R\sp
2$-terms,
and he mentioned that for $R+R_{,i}R_{,k}g\sp{ik}$ a
sixth-order field
equation appears. Remark: This lagrangian leads to the same
field
equation as \\$R-R\Box R$.

In [58], inflationary models with  a term $R_{,i} R\sp{,i}$ in
the action
are considered, but they do not vary with respect to the
metric, and so
no sixth-order term in the field equation appears.

Lu and Wise [59] consider the gravitational Lagrangian as a
sequence
$S=S_0+S_1+S_2+ \dots$ ordered with respect to physical
dimension.
So, $S_0=R$ and $S_1$ sums up the $R\sp 2$-terms. They try to
classify
the $S_2$-terms; however, their identity (8) is not correct,
so they
erroneously cancel the essential term $R\Box R$.

Kirsten et al. [60] consider the effective lagrangian for
self-interacting scalar fields;
in the renormalized action, the term
$$\frac{\Box R}{c+R}$$
appears. Wands [61] classifies lagrangians of the type $F(R,
\Phi)\Box R$ and
mentions that not all of them can be conformally
transformed to Einstein's
theory. Ref. [62] considers the lagrangian $\Phi \sp 2 \Box
R$, [63] the Lagrangian
$R \Box R$, [64, 65] double inflation from $\Phi $ and $R\sp
2$-terms,
also the $R\Box R$-terms is discussed.
Besides $R\Box R$ Berkin [65] considers the de Sitter
space-time as
attractor solution for field equations where the variational
derivative
of the term $C_{ijkl}C\sp{ijkl}$ is included.
The state of the art of the lagrangian $R\Box R$ can be found
in the papers [64-68]. \par
The paper is organized as follows: Sct. 2 compares several
possible definitions of an asymptotic de Sitter space--time,
sct. 2.1. for the set of spatially flat Friedman models, sct.
2.2. for less symmetric models. Sct. 3 deals with the
Lagrangian and corresponding field equations for higher--order
gravity. In sct. 4, we determine under which circumstances the
Bianchi models in higher--order gravity can be written in
diagonal form without loss of generality; the answer will be
more involved than the analogous problem for General
Relativity. In sct. 5 we discuss the results from the point of
view of structural stability in the sense
 of the ''Fragility''--paper [23].

\medskip

\section{Definitions \ of \ an \  asymptotic \ de Sitter
 space--time}
\setcounter{equation}{0}
In this section we want to compare some possible definitions
of an asymptotic de Sitter space--time.

\subsection{Spatially flat Friedman models}

Let us consider the metric
\begin{equation}
ds^2=dt^2 - e^{2\alpha(t)} \sum_{i=1}^n d(x^i)^2
\end{equation}
which can be called spatially flat Friedman model in $n$
spatial dimensions. We consider all values $n \ge 1$, but then

concentrate on the usual case $n=3$. If $n \le 3$ we often
write $x$, $y$, and $z$ instead of $x^1$, $x^2$, and $x^3$,
resp.  For metric (2.1) we define the Hubble parameter $H=
\dot \alpha \equiv \frac{d\alpha}{dt}$. We get
\begin{equation}
R_{00}= - n(\frac{dH}{dt} + H^2), \qquad
 R = - 2n(\frac{dH}{dt} + m H^2)
\end{equation}
where $m := \frac{n+1}{2} \ge 1$. We get
\newpage
\noindent {\bf Lemma 1}: The following conditions for metric
(2.1) are equivalent. \\
{\bf A}: It is flat. \  \  \  {\bf B}: $R=R_{00}=0$. \ \
\ {\bf C}: $R_{ij} R^{ij} =0$.\\
{\bf D}: $\alpha = const.$ or [$n=1$ and
$\alpha = \ln \vert t - t_0 \vert + const.$ ] \\
Proof: {\bf A} $\Rightarrow$ {\bf B} is trivial; {\bf B}
$\Rightarrow$ {\bf D} is done by solving the corresponding
differential equation; {\bf D} $\Rightarrow$ {\bf A} is
trivial for $\alpha = const$., the other case, i.e.,
$ds^2= dt^2 - (t-t_o)^2dx^2$, represents flat space--time in
polar coordinates; \\ {\bf C} $\Leftrightarrow$ {\bf B}
follows from the identity
\begin{equation}
R_{ij} R^{ij} = (R_{00})^2 + \frac{1}{n} (R-R_{00})^2
\end{equation}

An analogous statement can be formulated for the de Sitter
space--time. It holds \\
{\bf Lemma 2}: The following conditions for metric (2.1) are
equivalent. \\
{\bf A}: It is a non--flat space--time of constant
curvature.\\
{\bf B}: $R_{00}= \frac{R}{n+1}= const. \ne 0$. \\
{\bf C}: $(n+1) R_{ij} R^{ij} = R^2 = const. \ne 0  $.\\
{\bf D}: $H = const. \ne 0$ or [$n=1$ and
$ds^2= dt^2 - \sin^2(\lambda t)dx^2$
or $ds^2= dt^2 - \sinh^2(\lambda t)dx^2$  ] \\
The proof is analogous to lemma 1.

For $n=1$, the de Sitter space-time and anti-de Sitter
space-time differ by
the factor (-1) in front of the metric only. For $n>1$, under
the presumption of lemma 2, only the de Sitter space-time
($R<0$) is covered. Lemma 2 shows that within the class of
spatially flat Friedman models, a
characterization of the de Sitter space-time using polynomial
curvature
invariants only, is possible. \par
Next, let us look for isometries leaving the form of the
metric (2.1) invariant.
The function
\begin{equation}
\tilde \alpha(t) = c + \alpha ( \pm t +t_0), \qquad c, \ t_0 =
constants,
\end{equation}
 leads to an isometric space-time. The simplest expressions
being
invariant by such a transformation are $H^2$ and $\dot H$. We
take $\alpha $
as dimensionless, then $H$ is an inverse time and $\dot H$ an
inverse time squared.
Let $H \not= 0$ in the following. The expression
\begin{equation}
\varepsilon := \dot H H^{-2}
\end{equation}
is the simplest dimensionless quantity defined for the
spatially flat Friedman
models (2.1) and being invariant with respect to the
isometries (2.4). Let
$n>1$ in the following: Two metrics of type
(2.1) are isometric if and only
if the corresponding functions $\alpha $ and $\tilde \alpha $
are related
by equation (2.4). All dimensionless invariants containing at
most second
order derivatives of the metric can be expressed as $f(
\varepsilon )$,
where $f$
is any given function. But if one has no restriction to the
order, one gets
a sequence of further invariants
\begin{equation}
\varepsilon _2 = \ddot H H^{-3}, \ldots, \varepsilon _p =
                 \frac{d^pH}{dt^p} H^{-p-1}
\end{equation}
It holds: Metric (2.1) with $H\not= 0$ represents  the de
Sitter space-time
iff $\varepsilon \equiv 0$. \par
A third possible approach is the following: $\alpha (t) = Ht $

with $ H=const.
\not= 0$ is the de Sitter space--time, so we define an
asymptotic de Sitter space-time by the condition
\begin{equation}
\lim_{t\to \infty} \frac{\alpha(t)}{t} = const. \ne 0
\end{equation}
Let us summarize the variants Var(i) of the definitions.\\
{\bf Definition}: Let $H>0$ in metric (2.1) with $n>1$. We
call it an asymptotic de
Sitter space-time if \\
Var (1): $\lim _{t \to \infty} \frac{\alpha (t)}{t} = const.
>0$ \\
Var (2): $\lim _{t \to \infty} R^2 =const.>0$ and
         $\lim _{t \to \infty}(n+1)R_{ij}R{ij}-R^2=0$ \\
Var (3p): for $1\leq j\leq p$ it holds $\lim_{t \to \infty}
\varepsilon _p =0$.
All these definitions are different. One uses
$R_{ij}R^{ij}=n^2(\dot H + H^2)^2 + n (\dot H+nH^2)^2$.
However, as we will see in section 3.2, all these definitions
lead to the same
result if we restrict ourselves to the set of solutions of the
higher--order
field equations.

\subsection{Inhomogeneous cosmological models}

Let us start looking at the Kottler metric eq. (1.1). The
critical mass $M_{crit}= \frac{1}{3\sqrt{\Lambda}}$
 mentioned in subsection 1.3. in connection with the hoop
conjecture can be deduced (at least for the symmetries of
 the Kottler metric as follows: At a horizon, the function $A$

 must vanish.  One can see from eq. (1.1) that
for $\Lambda > 0$ the equation $A=0$ has solutions with
positive values $r$ if and only if $m \le M_{crit}$.
This means: the hoop conjecture is valid in the class of
spherically symmetric solutions. \par

However, this is not the problem we are dealing with here. The

problem is that none of the above definitions can be
generalized  to inhomogeneous models. One should find a
polynomial curvature  invariant which equals a positive
constant if and only if the  space--time is locally the de
Sitter space--time. To our  knowledge, such an invariant
cannot be found in the literature,  but also the
non--existence of such an invariant has not been   proven up
to now.

This situation is quite different for the positive definite
case: For signature $(++++)$ and
$S_{ij} = R_{ij}  - \frac{R}{4} g_{ij} $ it holds:
$$I \equiv  (R-R_0)^2 + C_{ijkl} C^{ijkl}  + S_{ij} S^{ij}
=0$$
iff the $V_4$ is a space of constant curvature $R_0$. So
$I \longrightarrow 0$ is a suitable definition of an
asymptotic  space of constant curvature.

One possibility exists, however, for the Lorentz signature
case,
 if one allows additional  structure as follows: An ideal
fluid has an energy--momentum  tensor
$$T_{ij} = ( \rho + p)u_i u_j - p g_{ij} $$
where $u_i$ is a continuous vector field with
$u_i u^i \equiv 1$. For stiff matter ($ \rho = - p$), the
equation
$T^{ij}_{\ \ ;j} \equiv 0$ implies $p= const.$, and so every
solution of  Einstein's theory with stiff matter is isometric
to a vacuum  solution of Einstein's theory with a cosmological
term. The  inverse statement, however, is valid only locally:

Given a  vacuum solution of Einstein's theory with a
$\Lambda $--term, one  has to find  continuous time--like unit
vector fields which need  not to exist from topological
reasons. And if they exist, they  are not at all unique.  So,
it becomes possible to define an  invariant $J$ which vanishes
iff the space--time is de Sitter by  transvecting the
curvature tensor with $u^i u^j$ and/or $g^{ij} $  and suitable
linear and quadratic  combinations of such terms.  Then time
$t$ becomes defined by the streamlines of the vector  $u^i$.
If one defines the asymptotic de Sitter space--time by
$J \longrightarrow 0$ as $t \longrightarrow \infty$, then it
turns out, that this definition is  {\it not} independent of
the vector field $u^i$.

\medskip

\section{ Lagrangian $F(R, \Box R, \Box \sp 2R,
\dots  ,  \Box \sp k R )$ }
\setcounter{equation}{0}

Let us consider the Lagrangian density $L$ given by
\begin{equation}
L = F(R, \Box R, \Box \sp 2R,
\dots  ,  \Box \sp k R) \sqrt{-g}
\end{equation}
where $R$ is the curvature scalar,  $\Box $ the D'Alembertian
and
$g_{ij}$  the metric of a (Pseudo-)Riemannian $V_D$ of
dimension
$D\ge   2$  and  arbitrary  signature;
 $g  =  - \vert  \det  \, g_{ij}  \vert  $ .
 The main application will be $D=4$ and  metric
signature $(+---)$.
 $F$ is supposed to be  a  sufficiently  smooth
function of its arguments, preferably a polynomial.
 Buchdahl  [50]  already  dealt with such kind of  Lagrangians

in 1951,  but then it became quiet of them for decades. Since
1990 a sequence  of papers on this topic appeared:  refs.
[51, 68]  for general $k$, and refs. [53 - 67] for the special
case $k=1$, i.e. the   Lagrangian  scalar is $F(R,  \Box R )$.

\subsection{The field equation}

The  variational  derivative of $L$ with respect  to  the
metric yields the tensor
\begin{equation}
P\sp{ij}  \  =  - \frac{1}{\sqrt{-g}} \
\frac{\delta  L  }{\delta g_{ij} }
\end{equation}
The components of this tensor  were given in the first paper
of ref. [51], their covariant components read
\begin{equation}
P_{ij} \ = \ G R_{ij} \ - \ \frac{1}{2} F g_{ij} \ - \ G_{;ij}
\ + \ g_{ij} \Box G \  + \ X_{ij}
\end{equation}
where the semi-colon denotes the covariant derivative,
 $R_{ij}$  the Ricci tensor, and
\begin{equation}
 X_{ij} \ = \ \sum_{A=1}\sp k \ \frac{1}{2} g_{ij}
[F_A(\Box\sp{A-1}  R)\sp{;m}  ]_{;m} \  - \
F_{A(;i}[\Box\sp{A-1}
R]_{;j)}
\end{equation}
having  the round symmetrization brackets in its last  term.
For $k=0$, i.e. $F = F(R)$, a case considered in sct. 4, the
tensor $  X_{ij}$  identically  vanishes.   It  remains  to
define  the expressions $F_A$, $A=0, \dots,k$ .
  The definition given in [51] can be simplified
as follows
\begin{equation}
F_k \ = \ \frac{\partial F}{\partial \Box \sp k R }
\end{equation}
and for $A=k-1, \dots,0$
\begin{equation}
F_A \ = \ \Box F_{A+1} \ + \
 \frac{\partial F}{\partial \Box \sp A R }
\end{equation}
and finally $G \ = \ F_0$.
The brackets are essential, for any scalar $\Phi $ it holds
\begin{equation}
\Box(\Phi _{;i}) \ - \ (\Box \Phi)_{;i}  \ = \ R_i \sp { \ j}
\ \Phi_{;j}
\end{equation}
Inserting  $\Phi  =  \Box\sp m R$ into this  equation,  one
gets
identities to be applied in the sequel without further notice.

It is well-known that
\begin{equation}
P\sp i _{\ j;i} \ \equiv \ 0
\end{equation}
and $P_{ij}$ identically vanishes  if and only
if $F$ is a divergence, i.e., locally there can be found a
vector
$v\sp i$ such that $F \ = \ v\sp i _{\ ;i}$ holds.  (Remark:
Even
for compact manifolds without boundary the restriction
''locally''
is unavoidable;  example:  Let $D=2$ and  $V_2$ be the
Riemannian
two-sphere $S\sp 2$ with arbitrary positive definite metric.
$R$
is  a divergence,  but there do not exist
 continuous
 vector fields $v\sp  i$
 fulfilling  $R \ = \ v\sp i _{\ ;i}$
 on the whole  $S\sp 2$.)

Example: for $m,n \ \ge \ 0$ it holds
\begin{equation}
\Box \sp m  R \ \Box \sp n R \ - \   R \ \Box \sp{m+n} R \ = \
divergence.
\end{equation}
So, the terms $\Box \sp m  R \ \Box \sp n R $ with naturals
$m$ and $n$ can be restricted to the case $m=0$ without loss
of generality. However, the more far--reaching statement by
Wands [61, page 271] ''Thus I can take any polynomial
 $F(\Box^i R)$ to be linear in its highest--order derivative
$\Box^n R$, multiplied by $F_n(R)$'' is not correct. Let us
give a counterexample: $R \Box R \Box R$, which leads to an
eighth--order field equation. \\
Proof that this is a counterexample: From dimensional reasons
only ingredients with $<length>^{-10}$ are to be considered.
Neglecting the divergencies, only the following ones are
candidates: $R^5$, $R^3 \Box R$, $R^2\Box^2R$,
$R\Box^3R$. They give rise to field equations of orders 4, 6,
8, and 10 resp. So the last term cannot be included. It
remains to look for
$$F= R \Box R \Box R + \gamma R^5 + \beta R^3 \Box R +
\alpha  R^2\Box^2R$$
The variation of $F$ with respect to the metric should vanish
identically. Vanishing of the 8th--order term requires
$\alpha = - \frac{1}{2}$. Vanishing of the 6th--order terms
gives rise to the equation
$$(\Box R + \frac{3\beta}{2}R^2)(\Box R)_{;ij} =0 $$
For no value of $\beta $ this is identically satisfied.

\subsection{Higher-order gravity}

We will examine the attractor property of the de Sitter
space-time in the
set of the spatially flat Friedman models. We need some useful
relations
for the de Sitter space-time:
\be						{24.5}
R=-n(n+1)H^2
\ee
and
\be						  {25}
R_{ij}  =  \frac{R}{n+1}g_{ij}\qmbox{and} \Box^k R  =  0
\qmbox{for} k>0~.
\ee
We insert this into the the field equation (3.3)
\be						  {27}
0  =  GR_{ij} - \half Fg_{ij}
   =  g_{ij}\li(\frac{1}{n+1}RG - \half F\re)
\ee
for the de Sitter space-time. The de Sitter space-time solves
the field
equation if and only if $2RG= D F$.
If we choose the lagrangian $(-R)^u$ with $u\in\R$ the
$D$-dimensional de
Sitter space-time satisfies the field equation iff
$\mbox{$u=\frac{n+1}{2}=\frac{D}{2}$}$.
We will examine the attractor property of the de Sitter
space-time in the set
of the Friedman models for the lagrangian $(-R)^u$ with
$\mbox{$2u=D=n+1>2$}$.
{}From this lagrangian it follows
\be						  {41}
F_A  =  0
\ee
and
\be					  {42}
G  =  -u(-R)~.
\ee
We get the field equation
\be					  {43}
0  =  -u(-R)^{u-1}R_{ij} - \half g_{ij}(-R)^u +
u\li[(-R)^{u-1}\re]_{;ij}
      - g_{ij}u\Box \li[(-R)^{u-1}\re] ~.
\ee
It is enough {}to examine the $00$-component of the field
equation, because all
the other components are fulfilled, if the $00$-component is fulfilled. We
make the ansatz
\be					  {44}
\dot\ga(t)  =  1 + \gb(t)
\ee
and get
\beq					  {45}
R_{00} & = & -n\dot\gb - 2n\gb - n \nonumber\\
R      & = & -2n\dot\gb -2(n^2 + n)\gb - (n^2 + n) \\
(-R)^m & = & 2nm(n^2 + n)^{m-1}\dot\gb + 2m(n^2 + n)^m\gb +
(n^2 + n)^m~.\nonumber
\eeq
One gets the field equation
\beq					  {46}
0\!\!\! & = &\!\!\! -2n^2u(u + 1)(n^2 + n)^{u-2}\ddot\gb
        - 2n^3u(u-1)(n^2 + n){u-2}\dot\gb + \nonumber\\
  &   &\!\!\! + nu(2u\! -\! n\! -\! 1)(n^2\!+ n)^{u-1}\gb
+\!\!\li(\!nu\! -\!
           \half(n^2\!+n)\!\re) \!\!(n^2\!+ n)^{u-1}.
\eeq
Using the condition $2u=n+1$ we get
\be					  {47}
0  =  \ddot\gb + n\dot\gb~.
\ee
All solutions of the linearized field equation are
\be					  {48}
\gb(t)  =  c_1 + c_2e^{-nt}~.
\ee
It follows
\be					  {49}
\ga(t)  =  t + \tilde c_1t + \tilde c_2e^{-nt} + \tilde c_3
\ee
and
\be					 {410}
\lim_{t\to\infty}\frac{\ga(t)}{t}  =  1+\tilde c_1~.
\ee
The $D$-dimensional de Sitter space-time is an attractor
solution
for the lagrangian $F=(-R)^{\frac{D}{2}}$.

The lagrangian $(-R)^{\frac{D}{2}}$ leads only {}to a field
equation of
fourth-order for $D>2$. The lagrangian $R\Box^k R$ with
$k>0$ gives a field equation of higher than fourth-order. For
this
case we get
\be					  {28}
F  =  R \Box^k R~,\qquad G  =  2 \Box^k R
\ee
and no solubility condition for $D>2$.
For the $00$-component of the field equation we need
\be					  {51}
F_A  =  \Box^{k-A}R
\ee
and
\be					  {52}
G  =  2\Box^kR
\ee
and get
\beq					  {53}
0 & = & \Box^kR\li(2R_{00}-\half R\re) + 2n\dot\ga\Box^kR_{,0}
+\nonumber\\
  &   & \hspace{1cm} + \sum_{A=1}^k(\Box^{k-A}R)(\Box^AR)
                     -
\half(\Box^{k-A}R)_{,0}(\Box^{A-1}R)_{,0}~.
\eeq
The ansatz (3.16)
leads {}to (3.17)
and
\be					  {56}
\Box(\Box^kR)  =  (\Box^kR)_{,00} + n(\Box^kR)_{,0}~.
\ee
We get the linearized field equation from (3.26)
\be					  {57}
\Box^kR  =  (\Box^kR)_{,0}~.
\ee
For $k=1$ we have
\be					  {58}
0  =  \gb^{(4)} + 2n\dddot{\gb} + (n^2 - n - 1)\ddot\gb +
(-n^2 - n)\dot\gb
\ee
with the characteristic polynomial
\be					  {59}
P(t)  =  x^4 + 2nx^3 + (n^2 - n - 1)x^2 + (-n^2 - n)x
\ee
possessing the roots $x_1=1$, $x_2=0$, $x_3=-n$ and
$x_4=-n-1$. We get the
solutions
\be					 {510}
\gb(t)  =  c_1 + c_2e^t + c_3e^{-nt} + c_4e^{-(n+1)t}
\ee
and
\be					 {511}
\ga(t)  =  \tilde c_1t + \tilde c_2e^t + \tilde c_3e^{-nt}
           + \tilde c_4e^{-(n+1)t}
\ee
and
\be					 {512}
\lim_{t\to\infty}\frac{\ga(t)}{t}  =  \infty~
\ee
unless $\tilde c_2=0$.
For the lagrangian $R\Box^k R$ the $D$-dimensional de Sitter
space-time is
not an attractor solution of the field equation. The formula
\be					 {513}
0  =  (\Box^{k+1}R)_{,0} - \Box^{k+1}R
   =  (\Box^kR_{,0} - \Box^kR)_{,00} + n((\Box^kR)_{,0} -
\Box^kR)_{,0}
\ee
for the linearized field equation for $k+1$ leads {}to the
recursive formula for
the characteristic polynomial:
\[
\ba{l}
\mbox{characteristic}\\ \mbox{polynomial for~~} k+1
\ea
=
\ba{l}
\mbox{characteristic}\\ \mbox{polynomial for~~} k
\ea
\cdot x \cdot (x+n)~.
\]
The characteristic polynomial for k has the roots:
\be					 {514}
\ba{lcll}
x_1 & = & ~~1  & \mbox{simple}\\
x_2 & = & ~~0  & \mbox{k-fold}\\
x_3 & = & -n   & \mbox{k-fold}\\
x_4 & = & -n-1 & \mbox{simple}~.\\
\ea
\ee
We get the solutions
\be					 {515}
\gb(t)  =  S(t) + T(t)e^{-nt} + c_1e^t + c_2e^{(-n-1)t}
\ee
and
\be					 {516}
\ga(t)  =  \widetilde S(t) + \widetilde T(t)e^{-nt} + \tilde
c_1e^t
           + \tilde c_2e^{(-n-1)t}
\ee
with $S,T,\widetilde T$ polynomials at most $k$-th degree and
$\widetilde S$
polynomial at most $k+1$-th degree. For most of all solutions
we get
\be					 {517}
\lim_{t\to\infty}\frac{\ga(t)}{t}  =  \infty
\ee
and therefore, the de Sitter space-time is not an attractor
solution for the
field equation derived from the lagrangian $R\Box^k R$.

These results have shown, that for the lagrangian
$R\Box^k R$ with $k>1$ the de Sitter space-time is not an
attractor
solution. The lagrangian $(-R)^{\frac{D}{2}}$ gives only a
fourth-order
differential equation. We will try {}to answer the following
question:\\
Are there generalized lagrangians so, that the de Sitter
space-time is an
attractor solution of the field equation?

First we make the ansatz
\be					  {61}
F  =  \sum_{k=1}^mc_kR\Box^kR \qmbox{with} c_m\neq0~.
\ee
In this case the de Sitter space-time is not an attractor
solution, because
for each term there one gets $+1$ as a root of the
characteristic polynomial
of the linearized field equation.

Now we make the ansatz
\be					  {62}
F  =  c_0(-R)^{\frac{D}{2}} + \sum_{k=1}^mc_kR\Box^kR
\qmbox{with} c_m\neq0~.
\ee
for the generalized lagrangian. One gets the characteristic
polynomial
\be					  {63}
\tilde P(x)  =  x(x+n)\li[c_0 + \sum_{k=1}^mc_kx^{k-1}(x +
n)^{k-1}(x - 1)(x + n + 1)\re]
\ee
for the linearized field equation. The solutions $x_1=0$ and
$x_2=-n$ do not
depend on the coefficients $c_i$ of the lagrangian. It is
sufficient {}to look
for the roots of the polynomial
\be					  {64}
P(x)  =  c_0 + \sum_{k=1}^mc_kx^{k-1}(x + n)^{k-1}(x - 1)(x +
n + 1)~.
\ee
If all solutions of this polynomial have negative real part,
then the
de Sitter space-time is an attractor solution for the field
equation. The
transformation
\be					  {65}
z  =  x^2 + nx + \frac{n^2}{4}
\ee
gives
\newpage
\beq					  {66}
P(x)\!\!
& = &\!\! Q(z) =\nonumber\\
& = &\!\! c_0 + \sum_{k=1}^mc_k\li(z - \frac{n^2}{4}\re)^{k-1}
          \li(z - \frac{n^2}{4} - n -1\re)\nonumber\\
& = &\!\! c_0 +
\sum_{k=1}^mc_k\!\!\li(\frac{n^2}{4}\re)^{k-1}\!\!
	  \li(\frac{n^2}{4} - n - 1\re)\!\! +
\!\!\sum_{l=1}^{m-1}\Bigg[c_l\!
	  + \!\sum_{k=l+1}^m
c_k\!\!\li(-\frac{n^2}{4}\re)^{k-l-1}\!\!\cdot\nonumber\\
&   &\!\! \cdot\!\!\li[{k-1 \choose
l-1}\!\!\li(-\frac{n^2}{4}\re)\!\! - \!\!
	  {k-1 \choose l}\!\!\li(\frac{n^2}{4} + n +
1\re)\re]\Bigg]z^l + c_m
	  z^m\nonumber\\
& = &\!\! d_0 + d_1z + \ldots + d_mz^m~.
\eeq
Now let
\be					  {67}
\ba{lcll}
a_{ll}\!\!\! & = & \!\!\!1  			& l = 0,\ldots,m\\
a_{0k}\!\!\! & = & \!\!\!-\li(-\frac{n^2}{4}\re)^{k-1}
		 \li(\frac{n^2}{4}+n+1\re)	& k = 1,\ldots,m\\
a_{lk}\!\!\! & = & \!\!\!\li(-\frac{n^2}{4}\re)^{k-l-1}\!\!
		 \li[{k-1 \choose l-1}\!\!\li(-\frac{n^2}{4}\re)\!\!
		 - \!\!\!{k-1 \choose l}\!\!\li(\frac{n^2}{4} + n +
1\re)\re]
					& l < k \le m\\
a_{kl}\!\!\! & = & \!\!\!0  			& \mbox{else}~.
\ea
\ee
This gives the equation
\be					  {68}
\li(\ba{c}
d_0 \\ \vdots \\ d_m
\ea\re)  =  A
\li(\ba{c}
c_0 \\ \vdots \\ c_m
\ea\re)\qmbox{with} A \quad\mbox{regular}~.
\ee
The roots of $P(x)$ have a negative real part iff the roots of
$Q(z)$ are from
the set
\be					  {69}
M :=  \li\{x+iy: x > \frac{n^2}{4} \wedge
|y|<n\sqrt{x-\frac{n^2}{4}}\re\}~.
\ee
If the roots $z_k$ of the polynomial $Q(z)$ are elements of
$M$, then the
coefficients $d_k$ are determined by
\be					 {610}
Q(z)  =  \sum_{k=0}^md_kz^k  = \prod_{k=1}^m(z-z_k).
\ee
The coefficients
\be					 {611}
\li(\ba{c}
c_0 \\ \vdots \\ c_m
\ea\re)  =  A^{-1}
\li(\ba{c}
d_0 \\ \vdots \\ d_m
\ea\re)
\ee
belong {}to a lagrangian, that gives a field equation with a
de Sitter
attractor solution.
The above considerations have shown that for every $m$ there
exists an
example for coefficients $c_k$, so that the de Sitter
space-time is an
attractor solution for the field equation derived from the
lagrangian
$c_0(-R)^{\frac{D}{2}} + \suml_{k=1}^mc_kR\Box^kR \qmbox{with}
c_m\neq0$.

It turned out that all the variants of the definition of an
asymptotic de Sitter solution given in subsection 2.1. lead to
the same class of solutions. \par
For the 6th--order case we can summarize as follows: \\
{\bf Theorem  1}: \ \ Let \ $L \ = \ R^2 \ + \ c_1 \ R  \
\Box R$ and
$L_E \ = \  R \ -  \ \frac{l^2}{6} \ L $ with length
$l>0$. Then the following statements are equivalent. \\
1. The Newtonian limit of $L_E$ is well--behaved, and the
potential $\phi$ consists of terms
$\frac{1}{r} e^{-\alpha r}$ with $\alpha \ge 0$ only.\\
2. The de Sitter space--time with $H=\frac{1}{l}$ is an
attractor solution for $L$ in the set of spatially flat
Friedman models, and this can already be seen from the
linearized field equation.\\
3. $c_1 \ge 0$ and the graceful exit problem is solved for the
quasi de Sitter phase $H\le 1/l$ of $L_E$.\\
4. $l^2= l_0^2+l_1^2$, $\quad  l^2 c_1 = l_0^2 l_1^2$ has a
solution with $0\le l_0<l_1$. \\
5. $0\le c_1 < \frac{l^2}{4}$. \\
For the proof of 3. one needs eq. (18) of the first of papers
[64] which reads in our notation $\dot H (1-4c_1H^2)=-1/6l^2$
showing that $\dot H < 0$ at the quasi de Sitter stage.\\
{\bf Theorem 2}: \  Let  $L$ and $L_E$ as in theorem 1.
  Then are equivalent: \\
1. The Newtonian limit of $L_E$ is well--behaved, for the
potential $\phi$ we allow $\frac{1}{r}$ and terms like
$\frac{P(r)}{r} e^{-\alpha r}$ with $\alpha > 0$ and a
polynomial $P$.\\
2. The de Sitter space--time with $H=\frac{1}{l}$ cannot be
ruled out to be an attractor solution for $L$ in the set of
spatially flat Friedman models if one considers the linearized
field equation only. \ \
3. $L_E$ is tachyonic--free.\\
4. $l^2= l_0^2+l_1^2$, $\quad l^2 c_1 = l_0^2 l_1^2$ has a
solution with $0\le l_0 \le l_1$. \ \
5. $0\le c_1 \le \frac{l^2}{4}$.

Of course, it would be interesting what happens in the region
where the linearized equation does not suffice to decide; one
should even not try to answer this question without a computer
algebra system.

\section{Higher-order gravity and diagonalizability of Bianchi
models}
\setcounter{equation}{0}
A Bianchi model can always be written as
\begin{equation}
ds\sp 2 = dt\sp 2 - g_{\alpha \beta} (t) \sigma \sp{\alpha}
\sigma \sp{\beta}
\end{equation}
where $g_{\alpha \beta}$ is positive definite and $\sigma
\sp{\alpha}$ are the characterizing one-forms. It holds
\begin{equation}
d\sigma \sp{\gamma} = - \frac{1}{2} C\sp{\gamma}_{\alpha
\beta} \sigma \sp{\alpha} \wedge \sigma \sp{\beta}
\end{equation}
with structure constants $C\sp{\gamma}_{\alpha \beta}$ of the
corresponding Bianchi type. It belongs to class A if
$C\sp{\beta}_{\alpha \beta} = 0$. The abelian group (Bianchi
type I) and the rotation group (Bianchi type IX) both belong
to class A.

In most cases, the $g_{\alpha \beta}$ are written in diagonal
form; it is a non-trivial problem to decide under which
circumstances this can be done without loss of generality.

For Einstein's theory, this problem is solved in [46]. One of
its results read:

If a Bianchi model of class A (except types I and II) has a
diagonal energy-momentum tensor, then the metric $g_{\alpha
\beta} (t)$ can be chosen in diagonal form. Here, the
energy-momentum tensor is called diagonal, if it is diagonal
in the basis $(dt,\ \sigma \sp 1,\ \sigma \sp 2,\ \sigma \sp
3)  $.

This result rests of course on Einstein's theory and cannot be
directly applied to higher-order gravity.

For fourth-order gravity following from a Lagrangian
$L=f(R)$ considered in an interval of $R$-values
where
$$\frac{df}{dR} \ \cdot \ \frac{d \sp 2f}{dR \sp 2} \ \ne \
0$$
one can do the following: The application of the conformal
equivalence theorem is possible, the conformal factor depends
on $t$ only, so the diagonal form of metric (3.1) does not
change. The conformal picture gives Einstein's theory with a
minimally coupled scalar field as source; the energy-momentum
tensor is automatically diagonal. So, in this class of
fourth-order theories of gravity, we can apply the above cited
theorem of MacCallum et al.

As example we formulate: All solutions of Bianchi type IX of
fourth-order gravity following from $L=R\sp 2$ considered in a
region where $R\ne 0$ can be written in diagonal form.

Consequently, the ansatz used in [27] by Barrow and
Sirousse-Zia
for this problem is already the most general one, cf. [30]
Spindel.

For fourth-order gravity of a more complicated structure,
however, things are more involved; example: Let
$$L=R+a R\sp 2 + b C_{ijkl} C\sp{ijkl}$$
with $ab \ne 0$. Then there exist Bianchi type IX models which
cannot be written in diagonal form. (This is a non-trivial
statement.)

To understand the difference between the cases $b = 0$ and $b
\ne 0$ it proves useful to perform the analysis independently
of the above cited papers [46]. For simplicity, we restrict to
Bianchi type I. Then the internal metric of the hypersurface
$[t=0]$ is flat and we can choose as initial value $g_{\alpha
\beta} (0)=\delta_{\alpha \beta} $. Spatial rotations do not
change this equation, and we can take advantage of them to
diagonalize the second fundamental form $\frac{d}{dt}g_{\alpha
\beta}(0)$.

First case: $b=0$. As additional initial conditions one has
only $R(0)$ and  $\frac{d}{dt}R(0)$. The field equation
ensures
$g_{\alpha \beta}(t)$
to remain diagonal for all times.

Second case: $b\ne 0$. Then one has further initial data
$\frac{d\sp 2}{dt\sp 2}g_{\alpha \beta}(0)$. In the generic
case, they cannot be brought to diagonal form simultaneously
with $\frac{d}{dt}g_{\alpha \beta}(0)$. This excludes a
diagonal form of the whole solution. (To complete the proof,
one has of course to check that these initial data are not in
contradiction to the constraint equations.) This case has the
following relation to the above cited theorem [46]: Just for
this case  $b\ne 0$, the conformal relation to Einstein's
theory breaks down, and if one tries to re-interpret the
variational derivative of $C_{ijkl}C\sp{ijkl}$ as
energy-momentum tensor then it turns out to be non-diagonal
generically, and the theorem cannot be applied.

For higher-order gravity, the situation becomes even more
involved. For a special class of theories, however, the
diagonalizability condition is exactly the same as in
Einstein's theory: If $L=R + \sum_{k=0}\sp m a_k R \Box \sp k
R$, ($a_m \ne 0)$ then in a region
  where $2L \ne R$
the Cauchy data are the data of General Relativity, $R(0)$,
and the first $2m+1$ temporal derivatives of $R$ at $t=0$. All
terms with the higher derivatives behave as an energy-momentum
tensor in diagonal form, and so the classical theorem applies.
[Let us comment on the restriction  $2L \ne R$ supposed above:
Eqs. (3.5, 3.6) show that $F_0 = G=0$ represents a singular
point of the differential equation (3.3); and for the
lagrangian given here $G=\frac{2L}{R} - 1$. For fourth--order
gravity defined by a non--linear lagrangian $L(R)$ one has
$G=\frac{dL}{dR}$ and $G=0$ defines the critical value of the
curvature scalar.]

\bigskip

\section{Structural stability of fourth-order cosmological
models}
In [23], Coley and Tavakol discuss cosmological models from
the point of view of structural stability; the notion for the
contrary of it is fragility. Structural stability is a more
general but less strictly defined notion than the usual
stability. So, its concrete meaning has context-dependently to
be specified.

1. Example: The Einstein universe (a closed Friedman model of
constant world radius in General Relativity with positive
cosmological term $\sim \Lambda$ and incoherent matter as
source) is unstable with respect to the initial data: A
non-vanishing but arbitrarily small initial Hubble parameter
gives rise to a singularity. This property ruled out the
Einstein universe as describing our real world. It should be
emphasized that this is in coincidence with the observational
result that our universe is not static, but that this
theoretical stability analysis  ruled out the Einstein
universe  independently of the observational result.

Structural stability represents stability not only with
respect to a small perturbation in the initial data, but a
small change in the corresponding type of matter and
 field equations. In most of the specifications one
 requires that by
a small change of conditions the qualitative (or topological)
properties of the system remain unchanged.
Concerning field equations, Coley and Tavakol [23] concentrate
on Lagrangians $L=f(R)$ for the gravitational field: For
linear functions $f$ one gets General Relativity, for
non-linear ones fourth-order gravity. Because of the change in
the order of the differential equation the question concerning
the robustness of General Relativity is a non-trivial one.
Before we follow this line we present some more or less
trivial examples from General Relativity to clearify what is
meant.

2. Example: The spatially flat Friedman model with incoherent
matter (dust) but $\Lambda = 0$ (Einstein-de Sitter  model)
has a scale factor $a \sim t\sp{2/3}$ for synchronized time
$t$. A small change of the initial data only changes the
proportionality factor, so this is stable. However, if we
consider this model within the class of all Friedman models,
then it represents just the bifurcating point between the
ever-expanding open and the recollapsing closed models. In
this sense, the Einstein-de Sitter model is a fragile one.

3. Example: Again we consider the Einstein-de Sitter model
within the class of all spatially flat Friedman models. We
impose new structure by allowing a new contribution to the
energy-momentum tensor in form of radiation not interacting
with the dust. During expansion, the energy density of the
radiation falls $\sim a\sp{-4}$, and of the dust only
$\sim a\sp{-3}$. The radiation becomes asymptotically
negligible, and, asymptotically for large values $t$, one gets
approximately $a \sim t\sp{2/3}$. In this sense, the
 Einstein-de Sitter model is structurally stable.

4. Example: Now we invert the point of view from the third
example. We start from a spatially flat Friedman filled with
radiation. Then one has $a \sim t\sp{1/2}$. We impose new
structure by adding an arbitrarily small amount of
non-interacting dust. As in the previous example, we get
asymptotically $a \sim t\sp{2/3}$. In this sense, the
spatially flat Friedman radiation model is structurally
unstable.

\bigskip

Let us now come to the consideration of Coley and Tavakol
concerning structural stability of fourth-order gravity
models. They consider perturbations of Friedman's radiation
model within fourth-order gravity. For the non-tachyonic case,
they get as result that the $R\sp 2$-term gives rise to an
instability. It is known for a long time, that asymptotically

the $R\sp 2$-term gives rise to damped oscillations which
behave as dust in the mean. So, the structural instability
considered there is exactly the same as in the 4. example
above and not a special feature of the fourth-order term.

Analogously they consider the quasi-de Sitter
stage (Starobinsky inflation) and get its stability for the
non-tachyonic case $L=R+aR\sp 2$.

\medskip

Remark: Coley, Tavakol [23] use the notion ''topological
almost all'' in the sense of ''countable intersection of open
dense subsets''. One must be careful in applying this notion,
especially, if one is tempted to mix it with the notion
''almost all'' in measure theory. A remarkable example shall
underline this warning: Let $I=[0,1]$ be the closed interval
with the usual probability measure $\mu $. Let
$\{r_n \vert n \in N\} \subset I$ be a countable dense subset
of $I$. For each natural $m$, let
$$A_m = I \cap \bigcup_{n\in N} ]r_n - 2\sp{-m-n} , r _n +
2\sp{-m-n}[ $$
and $A = \bigcap_{m\in N} A_m$.
(Here, $]x,y[$ denotes the open interval.)
Each $A_m$ is open and dense in $I$. For all values $m$,
$$\mu(A) \le \mu(A_m) \le \sum_{n\in N} 2\sp{1-m-n}
= 2\sp {1-m}$$
Hence, $\mu(A)=0$. So, $A$ contains topologically almost all
points of $I$, but there is zero probability to meet an
element of it.

Next, Coley and Tavakol consider structural stability of
Starobinsky inflation $L=R+aR\sp 2$ with respect to addition
of the cubic term $bR\sp 3$. For $L=R\sp 3$ alone one gets
polar inflation $a \sim t\sp{-10}$; considered in the region
$t<0$ this is expanding with $\dot h \, h\sp{-2} =
\frac{1}{10}$. The term with $b$ does not alter the order of
the differential equation, and so one expects a continuous
change of the properties. In fact, for small values $\vert b
\vert$ one has Starobinsky inflation as transient attractor,
with increasing  $\vert b \vert$  one gets a smaller basin of
attraction, and for $\vert b \vert \gg a\sp 2$ one needs
fine-tuned initial conditions.

A more drastic change of structure is to be expected if we
consider structural stability with respect to the addition of
terms like $R\Box R$.

\bigskip

\section{Discussion}
Sudarsky [66] proves the no--hair theorem (in the version that
there are no non--trivial black holes with regular horizon)
for the Einstein--Higgs theory. We have deduced a cosmic no
hair theorem on a quite different footing as follows (the more
detailed formulation is given at the end of sct. 3)
\newpage
\noindent
{\bf Theorem}: \ \ Let \ $L \ = \ R^2 \ + \ \epsilon \ l^2 \ R
\  \Box R$ and \\
$L_E \ = \ \frac{1}{16\pi G} \ [ R \ -  \ \frac{l^2}{6} \ L ]$
with length
$l>0$ and arbitrary real $\epsilon$. Then the following
statements are equivalent. \\
1. The Newtonian limit of $L_E$ is well--behaved.\\
2. The de Sitter space--time with $H=\frac{1}{l}$ is an
attractor solution for $L$.\\
3. $\epsilon \ge 0$ and the graceful exit problem is solved
for the quasi de Sitter phase $H\le 1/l$ of $L_E$.\\
4. $l^2= l_0^2+l_1^2$, $\quad \epsilon l^4 = l_0^2 l_1^2$ has
a solution with $0\le l_0<l_1$. \\
5. $0\le \epsilon < \frac{1}{4}$.

{}From the first glance this theorem is contrary to the
 results of refs.  [64 - 66]. But one should remember
that in refs. [64 - 66]  the question had been considered
whether the sixth--order terms   can typically lead to double
inflation.
The answer was: Double inflation (one period from
the $R^2$--term, the other one from the $R \Box R$--term)
requires a fine--tuning of initial conditions. Here we have
shown: The results of the Starobinsky model ($ \epsilon=0$ in
the present  notation) are structurally stable with respect to
the addition of  a sixth--order term $ \sim  \epsilon R \Box
R$, where $0 \le \epsilon < \frac{1}{4} $. The duration of the
transient quasi  de Sitter phase becomes reduced by a factor
$ \sim (1 - 4 \epsilon)$  only.

Further we have shown: For
$$L=R^2 + c_1 R \Box R + c_2 R \Box \Box R ,  \qquad c_2 \ne
0$$
and the usual case $n=3$ the de Sitter space--time with $H=1$
is  an attractor solution in the set of spatially flat
Friedman models if and only if the following inequalities are
fulfilled:
$$0<c_1< \frac{1}{4}, \qquad 0<c_2< \frac{1}{16}$$
and
$$c_1> - 13 c_2 + \sqrt{ 4c_2 + 225 c_2^2 } $$
This represents an open region in the $c_1-c_2$--plane whose
boundary contains the origin; and for the other boundary
points  the linearized equation does not suffice to decide the
attractor  property. This situation shall be called
''semi--attractor'' for  simplicity.  In contrary to the
 6th--order case, here we do not have  a one--to--one
correspondence, but a non--void open intersection with that
parameter set
having the Newtonian limit for $L_E$  well--behaved.

To find out, whether another de Sitter space--time with an
arbitrary Hubble parameter $H > 0$ is an attractor solution
for  the eighth--order field equation following from the
above  Lagrangian, one should remember that $H$ has the
physical  dimension of an inverted time, $c_1$ a time squared,
$c_2$ a time to  power 4. So, we have to replace $c_1$ by $c_1
H^2$ and
$c_2$ by $c_2 H^4$ in the above dimensionless inequalities to
get the correct conditions.
Example: $0<c_1 H^2 < \frac{1}{4}$.

\newpage

{\Large {\bf Acknowledgements}}

We thank W. Benz, U. Kasper, K. Peters, M. Rainer and S.
Reuter for valuable comments. Financial support from Deutsche
Forschungsgemeinschaft (contract Nr. Schm 911/6), from the
Wissenschaftler--Integrations--Programm (HJS) and from Potsdam
University (SK) are gratefully acknowledged.

\bigskip

{\Large {\bf References}}

\bigskip

[1] H. Weyl, Handbuch der Philosophie, chapter ''Philosophie
der Mathematik und Naturwissenschaft'' Oldenburg 1927.

[2] J. Barrow, G. G\"otz, Class. Quant. Grav. {\bf 6}, 1253
(1989).

[3] F. Hoyle, J. Narlikar, Proc. Royal Soc. (London)
 {\bf A 273}, 1 (1963).

[4] R. Price, Phys. Rev. {\bf D 5}, 2419 (1972);
 R. Price, Phys. Rev. {\bf D 5}, 2439 (1972).

[5] B. Altshuler, Class. Quant. Grav. {\bf 7}, 189  (1990).

[6] P. Peter, D. Polarski, A. Starobinsky, Phys. Rev. {\bf D
50}, 4827 (1994).

[7] U. Brauer, A. Rendall, O. Reula, Class. Quant. Grav. {\bf
11}, 2283 (1994).

[8] P. H\"ubner and J. Ehlers, Class. Quant. Grav. {\bf 8},
333 (1991); A. Burd, Class. Quant. Grav. {\bf 10}, 1495
(1993).

[9] G. Gibbons and S. Hawking, Phys. Rev. {\bf D 15}, 2738
(1977); S. Hawking and I. Moss, Phys. Lett. {\bf B 110}, 35
(1982);  M. Demianski, Nature {\bf 307}, 140 (1984).

[10] J. Barrow, Phys. Lett. {\bf B 180}, 335 (1986).

[11] K. Nakao, T. Shiromizu and K. Maeda, Class. Quant. Grav.
{\bf  11}, 2059 (1994).

[12] G. Murphy, Phys. Rev. {\bf D 8}, 4231 (1973); H. Oleak,
Astron. Nachr. {\bf 295}, 107 (1974).

[13] H. Oleak, Ann. Phys. (Leipz.) {\bf 44}, 74 (1987).

[14]  I. Prigogine, J. Geheniau, E. Gunzig and P. Nardone,
Proc. Natl. Acad. Sci. (USA) {\bf 85}, 7428 (1988);
I. Prigogine, J. Geheniau, E. Gunzig and P. Nardone, Gen.
Relativ. Grav. {\bf 21}, 767 (1989);
 I. Prigogine, J. Geheniau, E. Gunzig and P. Nardone,
Int. J. Theor. Phys. {\bf 28}, 927 (1989).

[15] U. Kasper, Int. J. Theor. Phys. {\bf 31}, 1007 (1992);
  U. Kasper, Acta Cosmologica {\bf 18}, 15 (1992).

[16] A. Vilenkin, Phys. Rev. {\bf D 46}, 2355 (1992);
 A. Borde and A. Vilenkin, Phys. Rev. Lett. {\bf 72}, 3305
(1994); A. Borde, Phys. Rev. {\bf D 50}, 3692 (1994).

[17] R. Mondaini, L. Vilar, Int. J. Mod. Phys. {\bf D 2}, 477
(1993).

[18] J. Pullin, Is there a connection between no-hair
behaviour and universality in gravitational collapse ?
 Preprint gr-qc/9409044, September 1994;
 C. Gundlach, R. Price and J. Pullin, Phys. Rev. {\bf D 49},
890 (1994); A. Strominger, L. Thorlacius, Phys. Rev. Lett.
{\bf 72}, 1584 (1994).

[19] T. Shiromizu, K. Nakao, H. Kodama and K. Maeda, Phys.
Rev. {\bf D 47}, R3099 (1993).

[20] M. Shibata, K. Nakao, T. Nakamura and K. Maeda, Phys.
Rev. {\bf D 50}, 708 (1994); T. Chiba, K. Maeda, Phys. Rev.
{\bf D 50}, 4903 (1994).

[21] S. Coleman, J. Preskill and F. Wilczek, Nucl. Phys. {\bf
B 378}, 175 (1992).

[22] J. Xu, L. Li, L. Liu, Phys. Rev. {\bf D 50}, 4886 (1994).

[23] A. Coley and R. Tavakol, Gen. Relat. Grav.
{\bf 24}, 835  (1992).

[24] H. Sirousse-Zia, Gen. Relat. Grav. {\bf 14}, 751 (1982).

[25] V. A. Belinsky, E. M. Lifshitz and I. M. Khalatnikov,
Sov. Phys. JETP {\bf 35}, 838 (1972).
Remark: The authors are in alphabetical order of the original
Russian language version.

[26] V. M\"uller, Ann. Phys. (Leipz.) {\bf 43}, 67 (1986).

[27] J. Barrow and H. Sirousse-Zia, Phys. Rev. {\bf D 39},
2187 (1989); Erratum {\bf D 41}, 1362 (1990).

[28] J. Yokoyama and K. Maeda, Phys. Rev. {\bf D 41}, 1047
(1990).

[29] S. Cotsakis, J. Demaret, Y. De Rop, L. Querella, Phys.
Rev. {\bf D 48}, 4595 (1993).

[30] P. Spindel, Int. J. Mod. Phys. {\bf D 3}, 273  (1994).

[31] B. Breizman, V. Gurovich and V. Sokolov, J. eksp. i teor.
Fiz. {\bf 59}, 288 (1970).

[32] J. Barrow, Phys. Lett. {\bf B 187}, 12 (1987);
 M. Pollock, Phys. Lett. {\bf B 192}, 59 (1987).

[33] V.  M\"uller, H.-J. Schmidt and A.A. Starobinsky, Phys.
Lett. {\bf B 202}, 198 (1988);  M. Mijic and J. Stein -
Schabes, Phys. Lett. {\bf B 203}, 353  (1988).

[34] K. Maeda, Phys. Rev. {\bf D 37}, 858  (1988);
 K. Maeda, Phys. Rev. {\bf D 39}, 3159  (1989);
 K. Maeda, J. Stein - Schabes and T. Futamase,
Phys. Rev. {\bf D 39}, 2848  (1989);
J. Barrow and P. Saich, Phys. Lett. {\bf  B 249}, 406  (1990).

[35] H. Feldman, Phys. Lett. {\bf B 249}, 200 (1990);
 B. Rogers and J. Isaacson, Nucl. Phys. {\bf B 364}, 381
(1991);  B. Rogers and J. Isaacson, Nucl. Phys. {\bf B 368},
415 (1992).

[36] J. Bi\v c\'ak, p. 12 in: Abstracts Conf. Gen. Relat.
13, Cordoba 1992; K. Maeda, ditto p. 296;  S. Cotsakis and G.
Flessas, Phys. Lett. {\bf B 319}, 69 (1993);  U. Kasper, S.
Kluske, M. Rainer, S. Reuter, H.-J. Schmidt, Stability
properties of the Starobinsky cosmological model, Preprint Uni
Potsdam 94/13, Oktober 1994,  gr-qc/9410030 SISSA Trieste;  H.
v. Borzeszkowski and H.-J. Treder, Found. Phys. {\bf 24}, 949
(1994).

[37] H.-J. Schmidt, Gen. Relat. Grav. {\bf 25}, 87  (1993);
Erratum p. 863.

[38] A. A. Starobinsky, Sov. Phys. JETP Lett. {\bf 37}, 66
(1983).

[39] A.A. Starobinsky and H.-J. Schmidt, Class. Quant. Grav.
{\bf 4}, 695  (1987).

[40] T. Pacher, On the no hair conjecture for inhomogeneous
space-times with a cosmological constant, Preprint 1986
Heidelberg. This preprint seems to be unpublished.

[41] L. Jensen and J. Stein-Schabes, Phys. Rev. {\bf D 35},
1146 (1987).

[42] M. Morris, Phys. Rev. {\bf D 39}, 1511 (1989).

[43] E. Calzetta and M. Sakellariadou, Phys. Rev. {\bf D 45},
2802 (1992).

[44] K. Nakao, T. Nakamura, K. Oohara and K. Maeda, Phys. Rev.
{\bf D 43}, 1788 (1991);
K. Nakao, K. Maeda, T. Nakamura and K. Oohara, Phys. Rev. {\bf
D 47}, 3194 (1993);
H. Shinkai and K. Maeda, Phys. Rev. {\bf D 48}, 3910 (1993);
H. Shinkai and K. Maeda, Phys. Rev. {\bf D 49}, 6367 (1994).

[45] A. Berkin, Phys. Rev. {\bf D 42}, 1016 (1990).

[46]  M. MacCallum, J. Stewart, B. Schmidt, Commun. Math.
Phys. {\bf 17}, 343 (1970); M. MacCallum, Phys. Lett. {\bf A
40}, 385 (1972).

[47] F. Kottler, Ann. Phys. (Leipz.) {\bf 56}, 410 (1918).

[48] P. Moniz, Phys. Rev. {\bf D 47}, 4315 (1993).

[49] H.-J. Schmidt, Fortschr. Phys. {\bf 41}, 179 (1993).

[50] H. Buchdahl, Acta Math. {\bf 85}, 63 (1951).

[51] H.-J. Schmidt, Class. Quant. Grav. {\bf 7}, 1023  (1990);
 I. Quandt and H.-J. Schmidt, Astron. Nachr. {\bf 312}, 97
(1991); S. Kluske and H.-J. Schmidt, Abstract 13.
\"Osterr. Math.-kongre\ss \ Linz Sept. 1993. 1 p;
 A. Battaglia Mayer and H.-J. Schmidt, Class. Quant. Grav.
{\bf 10}, 2441  (1993).

[52] C. Bollini, L. Oxman and M. Rocca,  J. Math. Phys. {\bf
35}, 4429 (1994).

[53] P. Forgacs, A. Wipf, J. Balog, L. Feher and L.
O'Raifeartaigh, Phys. Lett. {\bf B 227}, 214 (1989).

[54] G. Vilkovisky, Class. Quant. Grav. {\bf  9}, 895 (1992).

[55] G. Martin and F. Mazzitelli, Phys. Rev. {\bf D 50}, R613
(1994).

[56] K. Stelle, Phys. Rev. {\bf D 16}, 953 (1977).

[57] H.-J. Treder, Found. Phys. {\bf 21}, 283 (1991).

[58] S. Srivastava and K. Sinha, Phys. Lett. {\bf B 307}, 40
(1993).

[59] M. Lu and M. Wise, Phys. Rev. {\bf D 47}, R3095 (1993).

[60] K. Kirsten, G. Cognola and L. Vanzo, Phys. Rev. {\bf D
48}, 2813 (1993).

[61] D. Wands,  Class. Quant. Grav. {\bf 11}, 269  (1994).

[62] L. Amendola, Phys. Lett. {\bf B 301}, 175  (1993).

[63] S. Gottl\"ober, H.-J. Schmidt and A. A. Starobinsky,
 Class. Quant. Grav. {\bf 7}, 893  (1990).

[64] S. Gottl\"ober, V. M\"uller and H.-J. Schmidt,
 Astron. Nachr. {\bf 312}, 291  (1991);  S. Gottl\"ober, V.
M\"uller and A. A. Starobinsky, Phys. Rev. {\bf D 43}, 2510
(1991); H.-J. Schmidt, Proc. Sixth Marcel Grossmann Meeting on
General Relativity Kyoto, Eds.: H. Sato, T. Nakamura, WSPC
Singapore 1992, p. 92.

[65] A. Berkin and K. Maeda,  Phys. Lett. {\bf B 245}, 348
(1990);  A. Berkin and K. Maeda,
 Phys. Rev. {\bf D 44}, 1691  (1991); A. Berkin, Phys.  Rev.
{\bf D 44}, 1020  (1991).

[66] J. Hwang, Class. Quant. Grav. {\bf 8}, L133  (1991);
L. Amendola, S. Capozziello, M. Litterio and F. Occhionero,
Phys. Rev. {\bf D 45}, 417  (1992);
 S. Gottl\"ober, V. M\"uller, H.-J. Schmidt and A. A.
Starobinsky, Int. J. Mod. Phys. {\bf D 1}, 257  (1992);
 L. Amendola, A. Battaglia Mayer, S. Capozziello, S.
Gottl\"ober, V. M\"uller, F. Occhionero and H.-J. Schmidt,
 Class. Quant. Grav. {\bf 10}, L43  (1993);
D. Sudarsky,  Class. Quant. Grav. {\bf 12}, 579 (1995).

[67] H.-J. Schmidt, Phys. Rev. {\bf D 49}, 6354 (1994).

[68] S. Kluske, The de Sitter space-time as attractor solution
in higher order gravity, Preprint gr-qc/9501032, Potsdam Math
95/1, to appear in New Frontiers in gravitation Eds.: G.
Sardanashvily, R. Santilli, Hadronic Press 1995.

\end{document}